\documentclass{article}
\usepackage{spconf,amsmath,graphicx}
\usepackage{pifont}%
\newcommand{\cmark}{\ding{51}}%
\newcommand{\xmark}{\ding{55}}%
\usepackage[table,xcdraw]{xcolor}

\title{Noise robust speech emotion recognition with signal-to-noise ratio adapting speech enhancement}
\name{Yu-Wen Chen$^{\star}$ \qquad Julia Hirschberg$^{\star}$ \qquad Yu Tsao$^{\dagger}$}
  
\address{$^{\star}$ Department of Computer Science, Columbia University, United States\\
         $^{\dagger}$Research Center for Information Technology Innovation, Academia Sinica, Taiwan}
      
\begin{document}
%
\maketitle
\begin{abstract}
Speech emotion recognition (SER) often experiences reduced performance due to background noise. In addition, making a prediction on signals with only background noise could undermine user trust in the system. In this study, we propose a Noise Robust Speech Emotion Recognition system, NRSER. NRSER employs speech enhancement (SE) to effectively reduce the noise in input signals. Then, the signal-to-noise-ratio (SNR)-level detection structure and waveform reconstitution strategy are introduced to reduce the negative impact of SE on speech signals with no or little background noise. Our experimental results show that NRSER can effectively improve the noise robustness of the SER system, including preventing the system from making emotion recognition on signals consisting solely of background noise. Moreover, the proposed SNR-level detection structure can be used individually for tasks such as data selection.  

\end{abstract}
\section{Introduction}
\label{sec:intro}

Speech emotion recognition (SER) has many real-world applications, such as healthcare, stress monitoring, and marketing. However, in many use cases, users might not be speaking in a quiet environment. Therefore, the input signals are noisy, and a noisy input often leads to a significant drop in SER performance. Also, incorrect recognition of background noise may significantly reduce users' trust in the system; e.g., if a SER system recognizes different emotions while users change their environment without speaking, they might consider the system's recognition results unreliable. Hence, improving the noise robustness of SER is an essential research topic.

Previous studies have applied data augmentation to improve the noise robustness of SER. For example, \cite{xu2021head} contaminated noises to the speech signals, and \cite{tiwari2020multi} used a parametric generative model to generate the noisy data. In addition, researchers have incorporated several speech enhancement (SE) techniques, which aims to improve the quality and intelligibility of speech signals. For example, \cite{huang2013speech} introduced the spectral subtraction and masking based SE; \cite{chenchah2016speech} applied the spectral subtraction, wiener filter, and MMSE; \cite{ pohjalainen2016spectral} investigated SE methods in cepstral and log-spectral domains; and \cite{chakraborty2019front, triantafyllopoulos2019towards, zhou2020using} trained NN-based SE models. However, even though SE models can increase SER performance on noisy input speech signals, it also introduces artifacts and degrades SER performance on high signal-to-noise ratio (SNR) speech signals \cite{triantafyllopoulos2019towards}. Therefore, an automatic SNR-level detector is required to decide the suppression rule of SE \cite{xia2020weighted}. Recently, similar methods have been applied to SE on Automatic Speech Recognition (ASR) \cite{koizumi2021snri, sato2022learning}; however, the automatic SNR-level detector has not garnered sufficient attention in the integration of the SE model into SER.

In this study, we first investigate different strategies that apply SE to multi-task SER. Then, we propose a Noise Robust Speech Emotion Recognition (NRSER), which integrates the SE block, SNR-level detection block, and waveform reconstitution strategy. Our experimental results show that the proposed structure improves the system's performance on noisy speech signals without degrading its performance on signals with little or no background noise. Moreover, the proposed SNR-level detection structure can effectively distinguish between target noisy speech and signals that solely comprise background noise, which helps prevent the SER model from making predictions when the user is not speaking. Lastly, the SNR-level detection block can be used independently for other applications, such as data selection.

\section{Proposed NRSER}
\label{sec:proposed}
\begin{figure*}
\centerline{\includegraphics[scale=1.]{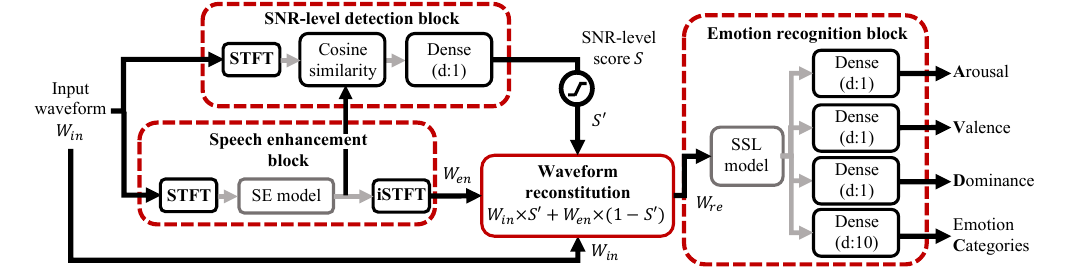}}
\caption{Proposed NRSER system, where $d$ in dense layers refers to the output dimension.}
\label{NRSER}
\end{figure*}

The NRSER system contains three NN blocks: SE, SNR-level detection, and emotion recognition (abbreviated below as "emotion block."). Figure~\ref{NRSER} depicts our NRSER system. 

\subsection{SE block}
In this study, we use the CMGAN\cite{cao2022cmgan} for SE. The CMGAN was pretrained using Voice Bank+DEMAND dataset and was fixed during the training of NRSER. Note that the CMGAN can be replaced with other SE models such as \cite{le21b_interspeech, fu2022uformer}. 

\subsection{SNR-level detection block}

We propose a simply yet effective SNR-level detection structure that can work with various SE models without requiring any modifications to the SE models. The idea behind the design of the SNR-level detection structure is that, for the high SNR-level signals, the enhanced signal should closely resemble the original signal since there is little to no noise to be removed. Conversely, for low-SNR signals with prominent background noise, the enhanced signal will be more different from the original signal due to the noise. Building on this idea, the SNR-level block calculates the similarity between the original and enhanced input signals, and then a dense layer predicts the SNR-level scores.

\subsection{Emotion block}
The self-supervised-learning (SSL)-based speech representation model aims to generate a general representation of the input signal. Previous studies have demonstrated the competitive generalizability and accessibility of using SSL models across various speech processing tasks, including SER\cite{wagner2023dawn}. We fine-tune and extend a pretrained SSL model into a multitask SER system. Specifically, a SSL model, HuBERT \cite{hsu2021hubert}, is finetuned by average-pooling the model's output embeddings and adding dense layers for arousal, valence, dominance, and emotion category recognition. In this study, we use \emph{hubert\_base} due to the limitation of computing resources. The \emph{hubert\_base} can be substituted with larger SSL models, such as \emph{hubert-large}, to achieve higher performance \cite{wagner2023dawn}. 

\subsection{NRSER structure}
First, an input waveform $W_{in}$ is fed to the SE block. Since the SE model we used is spectrogram-based, $W_{in}$ is converted to a spectrogram using the short-time Fourier transform (STFT). The output of the SE model is an enhanced spectrogram, which is used for the SNR-level detection and waveform reconstitution. The SNR-level detection block calculates the cosine similarity between the original and enhanced spectrograms in the time dimension. Then, a dense layer gives the SNR-level score $S$. The input waveform of the emotion block $W_{re}$ is reconstituted using the following equations: 
\begin{align}
S'= min(max(0, S), 1)
\end{align}
\begin{align}
W_{re}=W_{in}\times S'+W_{en}\times (1-S'),
\end{align}

where $W_{in}$ is the original input waveform, $W_{en}$ is the corresponding enhanced waveform, and $S'$ is the clamped SNR-level score. In the ideal condition, the $S'$ of a clean speech signal is 1, and the input of the emotion block $W_{re}$ is equal to the original input $W_{in}$. $W_{in}$ is used for emotion recognition without preprocessing since it does not contain background noise. In contrast, $S'$ of a noise signal is 0, and the input of emotion block $W_{re}$ is equal to the enhanced waveform $W_{en}$. $W_{in}$ is discarded because it only contains noise information that might mislead the emotion recognition.

\subsection{NRSER training phases}
The training of the NRSER system includes three phases. First, we train the SNR-level detection block. The training data contains clean speech and background noise signals, where the objective of speech and noise signals is set to 1 and 0, respectively. Then, we train the emotion block with speech signals processed by the SE block. Finally, we simultaneously fine-tune the emotion block with the reconstituted waveforms and the SNR-level detection block with the same data as the previous phase. 


\section{Experimental setup}
\label{sec:exp_setup}

\subsection{Data}

For emotion recognition, we adhere to the partitions defined in the MSP-PODCAST v1.10 dataset \cite{lotfian2017building} on the training, validation, and testing sets. The MSP-train, MSP-val, and MSP-test represent the train, development, and collection of test1 and test2 sets. The number of segments for MSP-train, MSP-val, and MSP-test is 63,076, 10,999, and 16,903+13,289, respectively. The noisy utterances for training and validation were created by contaminating utterances in MSP-train and MSP-val (signals) with one random sample from the balanced training set of AudioSet (noise) dataset \cite{gemmeke2017audio} at three SNR levels (6 dB, 10 dB, and 14 dB). In total, the training and validation noises contain 15,651 and 3913 segments, respectively. Finally, the utterances in MSP-test (signals) were contaminated with one random sample from the validation set of AudioSet (noise) dataset at 8 dB and 12 dB. In total, validation set of AudioSet dataset (denoted as AudioSet-val) includes 17,903 noise segments. For the training and validation of the SNR-level detection block, we use MSP-train and MSP-val as clean speech and balanced training set of AudioSet (denoted as AudioSet-train) as noise. 


\subsection{Training and system hyperparameter}
The sampling rate of all signals were set to 16kHz. The parameter setup of the STFT was a Hanning window with a window length of 400 and the hop length of 100. For the SE block, we used the open-sourced pretrained CMGAN\cite{cao2022cmgan}. The weight of the SE block was fixed during the training and the inference steps. The SNR-level detection block was trained using a batch size of 32, an SGD optimizer with a learning rate of 0.0001 and a momentum of 0.9, and the mean square error (MSE) loss criterion. The emotion block was trained using a batch size of 8, an SGD optimizer with a learning rate of 0.0001 and a momentum of 0.9. The training loss of emotion block is a combination of the cross-entropy loss for the emotion category recognition and the concordance correlation coefficient (CCC) loss \cite{li2021contrastive} for arousal, valence, and dominance recognition. The emotion category consists of ten primary emotions categories as defined in MSP-PODCAST \cite{lotfian2017building}. All models were trained using early stopping, with a patience of 2. The weights corresponding to the best validation scores during training were saved. The experimental results were the average of the same model and data retrained with different random seeds three times. 

\section{Results}
\label{sec:results}

\subsection{Performance on speech emotion recognition}

We compared NRSER performance with four representative configurations of the SER system (S-clean, S-noisy, S-en, and S-en'.) S-clean is the baseline system that only contains the emotion block and is trained with the MSP. S-noisy system is the S-clean system trained with additional MSP+AudioSet training data. S-en, S-en', and NRSER all incorporate SE into SER. S-en is S-clean using the SE as preprocessing in inference. Compared with S-en, the S-en' system uses the SE in both training and testing time. Table~\ref{table:system_configuration} summarized the configurations of SER systems.

\begin{table}
\caption{SER system configurations}
\centering
\resizebox{\columnwidth}{!}
{%
\begin{tabular}{|c|c|c|c|c|c|}
\hline
\rowcolor[HTML]{EFEFEF} 
                 & S-clean & S-noisy & S-en & S-en' & NRSER \\ \hline
Emotion block    & \cmark       & \cmark       & \cmark    & \cmark     & \cmark     \\ \hline
SNR block     & \xmark       & \xmark       & \xmark    & \xmark     & \cmark     \\ \hline
SE block (Train) & \xmark       & \xmark       & \xmark    & \cmark     & \cmark     \\ \hline
SE block (Test)  & \xmark       & \xmark       & \cmark    & \cmark     & \cmark     \\ \hline
Training data &
  MSP &
  \begin{tabular}[c]{@{}c@{}}MSP,\\ MSP+AudioSet\end{tabular} &
  MPS &
  \begin{tabular}[c]{@{}c@{}}MSP,\\ MSP+AudioSet\end{tabular} &
  \begin{tabular}[c]{@{}c@{}}MSP,\\ MSP+AudioSet\\AudioSet\end{tabular} 
  \\ \hline
\end{tabular}%
}
\label{table:system_configuration}
\end{table}

\begin{figure}[htbp!]
\centerline{\includegraphics[scale=0.98]{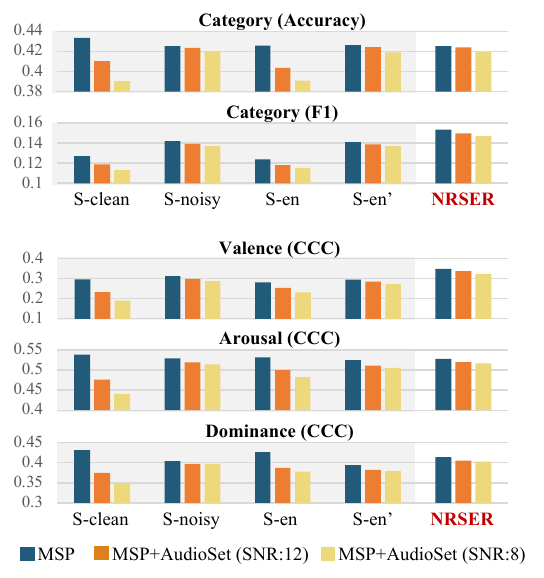}}
\caption{SER performance.}
\label{fig:system_performance}
\end{figure}

To evaluate the noise robustness of the systems, we test the systems on data with different noisy levels: MSP-PODCAST data (MSP), and MSP mixed with AudioSet-val noise (MSP+AudioSet (SNR:12) and MSP+AudioSet (SNR:8).) The results are presented in Figure~\ref{fig:system_performance}. All systems performed worse with increasing noise levels (i.e., MSP $>$ MSP+AudioSet (SNR:12) $>$ MSP+AudioSet (SNR:8)). As the input signals became noisier, the performance of S-clean and S-en decreased significantly, while S-noisy, S-en', and NRSER exhibited greater noise robustness. S-clean generally performed better than S-en on MSP but worse than S-en on MSP+AudioSet, revealing that SE degrades SER performance on signals with no or little background noise. S-en' outperformed S-en in most cases since it was trained using enhanced signals. Because NRSER mitigated the effects of distortion caused by SE, it achieves better performance than S-en'. Overall, we observed that NRSER was less effective in recognizing arousal and dominance. One possible reason is that the SE process reduces the signal intensity while removing noise, potentially leading to the loss of information related to arousal and dominance. However, NRSER provided more specific information about the emotional experience, achieving the best performance in terms of F1 score for emotion category recognition and valence. 


\subsection{Analysis of the SNR-level detection block}
\label{vad_with_snr}

\begin{figure}[htb]
\centerline{\includegraphics[scale=1.]{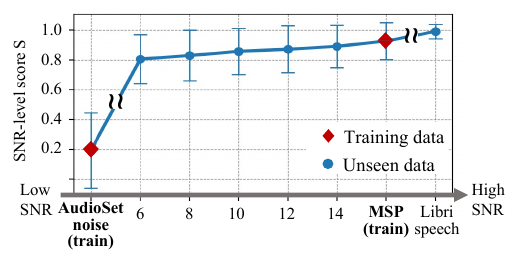}}
\caption{Average SNR-level score $S$ on signals with different SNR levels.}
\label{fig:snr_analysis}
\end{figure}

To verify the effectiveness of the SNR-level detection block, we calculated the average SNR-level score $S$ of signals at different SNR levels (Figure~\ref{fig:snr_analysis}). The red diamonds indicate the average SNR-level score $S$ of the training data, including the AudioSet-train and MSP-train. During training, MSP-train was labeled as $1$, whereas AudioSet-train was labeled as $0$. The blue circles show the average $S$ of unseen data for the tested SNR-level detection model. The numbers refer to the MSP (signal) to AudioSet (noise) ratio level for the MSP+AudioSet data. The results reflect that the SNR-level detection block can recognize the interval levels even though it was trained using only the upper and lower bounds. Moreover, we tested the SNR-level detection block on the LibriSpeech test-clean set \cite{panayotov2015librispeech}, in which the recordings have higher quality than the training data (MSP-train.) The result indicates that the SNR-level detection block can recognize that LibriSpeech signals have a higher SNR level than the MSP-train. Notably, this outcome was achieved even though all signals in MSP-train data were assigned the maximum objective score (i.e., 1) during training. 

We also analyzed the SNR-level score in distinguishing noisy speech (MSP-test mixed with AudioSet-val data) and background noise signals (AudioSet-val). The experimental results show that setting threshold of 0.6 achieves more than 90\% accuracy in our case. By using the SNR-level score as an indicator, NRSER can avoid attempting to provide emotion recognition results from signals with only background noise. Note that such attempts give users a clear sign of a system error and might diminish their trust in the system. In addition, the SNR-level detection block can be used in other speech-processing tasks. For example, it can aid data selection by filtering out low-quality data from large crawled datasets.

\subsection{Effectiveness of the waveform reconstitution}

\begin{figure}[htb]
\centerline{\includegraphics[scale=0.9]{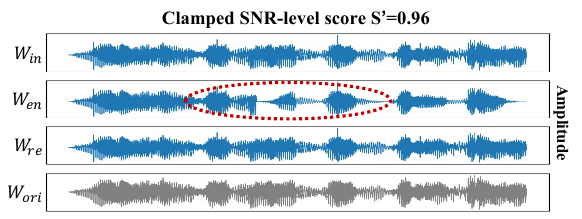}}
\caption{Visualization of the speech signals. The red circle highlights the distortion caused by the SE. Note that this figure shows the segments rather than the entire signals.}
\label{fig:waveform}
\end{figure}

We visualize the effectiveness of waveform reconstitution on signals with relatively higher clamped SNR-level score $S'$ in Figure~\ref{fig:waveform}, where $W_{ori}$ is an utterance in MSP-test, and $W_{in}$ is the utterance contaminated with additional noise (MSP+AudioSet). Because $S'$ was calculated based on the similarity between the spectrogram with and without being processed by the SE block, an input signal receives a higher $S'$ if the SE block makes less change to the input signal. There are two possible cases where an input signal received a high $S'$: (1) the input signal does not have background noise, and thus the SE block does not need to remove the noise; (2) the SE block could not remove the noise from the input signal. Figure~\ref{fig:snr_analysis} has already revealed case (1), that signals with higher SNR levels receive higher $S'$. Figure~\ref{fig:waveform} demonstrates case (2), that $W_{in}$ has a high $S'$ because the SE block fails to remove the noise. In both cases, the SER system did not benefit from the SE structure because the SE did not remove the noise but caused the distortion. However, since the SNR-level score is high, the reconstituted waveform $W_{re}$ preserves the speech information in the $W_{in}$ and reduces the distortion caused by the SE structure. 

\section{Conclusion}
\label{sec:conclusion}

We propose NRSER, a noise-robust SER system. Our experimental results confirm the effectiveness of integrating the SE block, SNR-level detection block, and waveform reconstitution strategy in improving the noise robustness of the SER system. The experimental results also demonstrate that training the SNR-level block with the upper and lower bounds of the objectives is sufficient for the model to learn the interval levels and recognize data that falls outside the boundary. Notably, the SE and emotion blocks can be substituted by alternatives that exhibit superior performance compared to the ones employed in this study, thereby achieving higher overall performance. Finally, we hope that our proposed noise-robust structure is not limited to emotion recognition but can also be applied to improve the noise robustness of other speech-processing tasks.



\vfill\pagebreak

\bibliographystyle{IEEEbib}
\bibliography{strings,refs}

\end{document}